\documentclass[a4papar]{article}
\usepackage{amsmath}
\usepackage{amssymb}
\usepackage{graphicx}
\usepackage{cite}
\usepackage{authblk} 

\newcommand{\av}[1]{\langle #1 \rangle}
\newcommand{\avnu}[1]{\langle #1 \rangle^{(\nu)}}
\newcommand{\avnuq}[2]{\langle #1 \rangle^{(\nu=#2)}}
\newcommand{\nudelta}[1]{#1_{\delta}^{(\nu)}}
\newcommand{\nudeltac}[1]{\left( #1_{\mathrm{c}} \right)_{\delta}^{(\nu)}}

\title{Transverse momentum fluctuation under the Tsallis distribution at high energies}
\author{Masamichi Ishihara}
\affil{Department of Human Life Studies,\\
 Koriyama Women's University,\\
Koriyama, Fukushima, 963-8503, Japan\\
}

\date{}

\begin{document}
\maketitle

\begin{abstract}
We studied the effects of the Tsallis distribution on the transverse momentum fluctuation in high energy collisions. 
The parton-hadron duality and the Bose-Einstein type correlation between partons were assumed. 
The fluctuation was calculated in the boost-invariant picture 
for the expectation value used in the Boltzmann-Gibbs statistics and 
for the expectation value used in the Tsallis nonextensive statistics.
It was shown that the fluctuation is a function of $\eta$ which is the ratio of the inverse temperature to the correlation length.
We found the following points: 
(1) the fluctuation depends on the form of the distribution and depends weakly on 
the definition of the expectation value used in the statistics,
(2) the fluctuation increases as the entropic parameter value of the Tsallis distribution increases, 
and 
(3) the variation of the fluctuation as a function of the entropic parameter 
for the expectation value used in the Boltzmann-Gibbs statistics is larger than 
that for the expectation value used in the Tsallis nonextensive statistics in the wide range of $\eta$.
\end{abstract}


\section{Introduction}

Power-like distributions are of interest in various branches of physics,  
and 
a momentum distribution of produced particles in a high energy collision shows a power-like distribution
\cite{Alberico2000, Wilk2007, Osada2008, Alberico2009, Cleymans2012, Marques2015, Cleymans2017, GS2014, GS2015, Zheng2016, Lao2016, our}.
One of power-like distributions is a Tsallis distribution which has two parameters: 
the temperature and the entropic parameter $q$.
It was reported that the (momentum) distribution of the produced particles in a high energy collision 
is described well by a Tsallis distribution. 
Therefore, the Tsallis distribution has been of interest at high energies. 

The distribution affects physical quantities, such as fluctuation.
A Tsallis-type function in the Geometrical Scaling picture describes well the distribution of emitted particles\cite{our}. 
Another Tsallis-type function has been used to describe the distribution \cite{Cleymans2012, Marques2015, Cleymans2017}.
Finding the function which describes the distribution well is important, 
because the distribution is useful to understand the processes in the collisions.

The value of a physical quantity depends on the definition of the expectation value used in the statistics.
The Boltzmann-Gibbs (BG) statistics is generally applied to thermally equilibrated systems. 
In the past few decades, the Tsallis nonextensive statistics has been applied to various systems in which a power-like distribution appears. 
The (conventional) expectation value used in the BG statistics is different from 
the expectation value used in the Tsallis nonextensive statistics \cite{tsallis1998, lavagno2002}.
The differences were found in chiral phase transitions \cite{Ishihara2015:BG, Ishihara2016:Tsallis}.
Physical quantities depend on the definition of the expectation value used in the statistics, 
and therefore the effects of the difference in the definition of the expectation value on the quantities should be studied.

A fluctuation between produced particles is related to the physical picture, such as glasma \cite{Lappi:conf:2011}. 
As discussed above, the value of the fluctuation depends on 
the form of the distribution and the definition of the expectation value.
Therefore, it is necessary to study how the fluctuation depends on the form of the distribution 
and the definition of the expectation value in order to obtain the information from the fluctuation. 

The emitted hadrons come from the generated particles, partons, in high energy collisions. 
It is plausible that these hadrons take over the momentum distribution of the partons (parton-hadron duality) .
The concept of the duality appeared in the study of the scaling behavior of hadrons \cite{Bloom1971}.
The duality was also used to study the physical quantities, such as structure functions \cite{Steffens2004} and correlators \cite{Boris1996}.
In the same way, it is natural that these hadrons take over the fluctuation of the partons.
The hadrons generated from the partons distribute and correlate at high energies.  

In this paper, we study the fluctuation of hadrons at high energies.  
The purposes are 
(i) obtaining the fluctuation of hadrons when the distribution of partons is given by a Tsallis distribution, 
(ii) showing the differences between the fluctuation for the expectation value used in the BG statistics 
and the fluctuation for the expectation value used in the Tsallis nonextensive statistics,  
and 
(iii) showing the differences in the fluctuation 
between the distribution appearing in the Geometrical Scaling picture \cite{Kharzeev,GS2014,GS2015}
and the distribution based on Tsallis entropy.

We summarize briefly the procedure and the results. 
We assume parton-hadron duality in a one-dimensional expanding system:
the fluctuation between hadrons reflects directly the fluctuation between partons.
A Bose-Einstein type correlation is introduced as a correlation between partons.
We found the following points:
(1) the fluctuation as a function of the ratio of the inverse temperature to the correlation length 
depends on the form of distribution, 
and depends weakly on the definition of the expectation value used in the statistics, 
(2) the fluctuation increases as $q$ increases, 
and 
(3) the variation of the fluctuation as a function of $q$ for the expectation value used in the BG statistics is larger than 
that for the expectation value used in the Tsallis nonextensive statistics
in the wide range of $\eta$, where $\eta$ is the ratio of the inverse temperature to the correlation length.

This paper is organized as follows. 
In section~\ref{sec:correlation}, 
we introduce the momentum distribution used in this study.
We calculate the average of the absolute value of the transverse momentum 
for the expectation value used in the BG statistics and 
that for the expectation value used in the Tsallis nonextensive statistics. 
We also calculate the fluctuation for these expectation values. 
In section~\ref{sec:numerical}, the numerical results are shown. 
The fluctuations are calculated numerically as functions of $\eta$ for various $q$.
The last section is assigned for discussion and conclusion.

\section{Transverse Momentum Fluctuation}
\label{sec:correlation}
\subsection{Momentum Distribution Function}
We introduce the momentum distribution function in this subsection,  
because the physical quantities are calculated 
under the momentum distribution for the expectation value used in the BG statistics and 
for the expectation value used in the Tsallis nonextensive statistics. 
The parameter $\nu$ is used to distinguish the expectation values. 
The quantity $\nu$ is $1$ for the expectation value used in the BG statistics 
and $q$ for the expectation value used in the Tsallis nonextensive statistics. 
The transverse momentum $\vec{p}_T$ and rapidity $y$ are used as variables, 
and the absolute value of $\vec{p}_T$ is represented as  $p_T$. 
The number of partons and that of charged hadrons are represented as $N^{(\nu)}$ and  $N_{\mathrm{c}}^{(\nu)}$ respectively. 
We add the subscript $\mathrm{c}$ to the quantities for charged hadrons.
The parton-hadron duality is assumed:
\begin{equation}
\frac{d^3N_{\mathrm{c}}^{(\nu)}}{d\vec{p}_T^2 dy}   = \frac{1}{\gamma} \left( \frac{d^3N^{(\nu)}}{d\vec{p}_T^2 dy} \right) , 
\end{equation}
where $\gamma$ is constant. 
Therefore, we can calculate some quantities without $N_{\mathrm{c}}^{(\nu)}$.
We use the quantity $\frac{d^2N^{(\nu)}}{d\vec{p}_T^2}$:
\begin{equation}
\frac{d^2N^{(\nu)}}{d\vec{p}_T^2} = \int dy \left( \frac{d^3N^{(\nu)}}{d\vec{p}_T^2 dy} \right) . 
\label{eqn:basic-y-integ}
\end{equation}
When the $\frac{d^3N^{(\nu)}}{d\vec{p}_T^2 dy}$ is approximately independent of $y$ (boost-invariant), 
we use the following integral with respect to $y$, referring the value at $y=0$:
\begin{equation}
\frac{d^2N^{(\nu)}}{d\vec{p}_T^2} = \int_{\Delta y} dy \left. \frac{d^3N^{(\nu)}}{d\vec{p}_T^2 dy} \right|_{y=0} . 
\label{eqn:bjorken}
\end{equation}

The momentum distribution is a basic quantity in high energy collisions. 
We give the general form of the distribution, 
though we assume massless classical particles with vanishing chemical potential.
As a choice, we postulate the following form which appears in the Geometrical Scaling picture \cite{GS2015} 
in the studies of high energy collisions: 
\begin{equation}
\frac{d^2N^{(\nu)}}{d\vec{p}_T^2} = \mathrm{Constant} \times \left( \frac{1}{e_q\left( \beta (p_T - \mu) \right) +\xi } \right)^{\nu} , 
\label{eqn:dist-form-GStype}
\end{equation}
where $\mu$ is chemical potential, $\beta$ is the inverse of the temperature. 
The quantity $\xi$ is $-1$ for boson, $0$ for classical particle, and $1$ for fermion. 
The function $e_q(x)$ is the q-exponential function defined below.
We note that both $\mu$ and $\xi$ are zero in this study.
As another choice, we use the following form \cite{Cleymans2012, Marques2015, Cleymans2017} for massless particles,
which is derived from the Tsallis entropy:
\begin{equation}
\frac{d^2N^{(\nu)}}{d\vec{p}_T^2} = \mathrm{Constant} \times p_T \left( \frac{1}{e_q\left( \beta (p_T - \mu) \right) +\xi } \right)^{\nu} . 
\end{equation}
Therefore, we use the following form as the transverse momentum distribution function:
\begin{equation} 
\frac{d^2N^{(\nu)}_{\delta}}{d\vec{p}_T^2} = \frac{\nudelta{C}}{(2\pi)} \left(p_T\right)^{\delta} \left( \frac{1}{e_q\left( \beta (p_T - \mu) \right) +\xi } \right)^{\nu} , 
\label{def:momentumdist:universal}
\end{equation}
where the two cases are distinguished with $\delta$: $\delta = 0, 1$. 
The quantity $\nudelta{C}$ is constant.
The function $e_q(x)$ is given\cite{Cleymans2012, Conroy2010} as 
\begin{equation}
e_q(x) := \left\{
\begin{array}{ll}
\left[1+(q-1) x \right]^{1/(q-1)} & \qquad ( x \ge 0 )\\
\left[1+(1-q) x \right]^{1/(1-q)} & \qquad ( x < 0 )
\end{array}
\right. 
.
\end{equation}
We deal with massless classical particles ($\xi=0$) with $\mu=0$ in the present study. 
We have
\begin{equation} 
\frac{d^2N^{(\nu)}_{\delta}}{d\vec{p}_T^2} 
= \frac{\nudelta{C}}{(2\pi)} \left(p_T\right)^{\delta} \left( \frac{1}{e_q\left( \beta p_T \right) } \right)^{\nu} ,  \qquad 
(\nu=1,q; \quad \delta =0, 1)
.
\label{eqn:used-dist}
\end{equation}
These distributions were also found in the studies of the distributions of the emitted particles at high energies \cite{Zheng2016,Lao2016}.
We calculate the average of $p_T \equiv |\vec{p}_T|$ and the transverse momentum fluctuation 
with Eq.~\eqref{eqn:used-dist} in the following subsections.

\subsection{The average of the absolute value of the transverse momentum under the Tsallis distribution}
We now focus on the average of $p_T$ of the emitted particles in high energy collisions.
We deal with the distributions of $q>1$, 
because it is reported that the entropic parameter $q$ is larger than $1$ in high energy collisions
\cite{Alberico2000, Wilk2007, Osada2008,  Alberico2009, Cleymans2012, Marques2015, Cleymans2017, our}
.

The average of $p_T$ of charged hadrons is given by 
\begin{equation}
\avnu{p_T}_{\mathrm{c}} = \avnu{p_T} = 
\frac{\displaystyle \int d^2\vec{p}_T \int dy p_T \left( \frac{d^3N^{(\nu)}}{d\vec{p}_T^2 dy} \right)}
{\displaystyle \int d^2\vec{p}_T \int dy \left( \frac{d^3N^{(\nu)}}{d\vec{p}_T^2 dy} \right)}
, 
\end{equation}
when the parton-hadron duality is a good approximation.
This average with the distribution, Eq.~\eqref{eqn:used-dist}, for massless classical particles with vanishing chemical potential is 
\begin{equation}
\avnu{p_T}_{\delta} (\beta; q) = 
\frac{\displaystyle \int dp_T  \left( p_T \right)^{2+\delta} \left( \frac{1}{e_q\left(\beta p_T\right)} \right)^{\nu}}
{\displaystyle \int dp_T  \left( p_T \right)^{1+\delta} \left( \frac{1}{e_q\left(\beta p_T\right)} \right)^{\nu}}
,
\end{equation}
where we attach the subscript $\delta$ to distinguish the distributions.
In the present situation, 
the average for the expectation value used in the Tsallis nonextensive statistics $\avnuq{p_T}{q}_{\delta}$ and 
the average for the expectation value used in the BG statistics $\avnuq{p_T}{1}_{\delta}$ are related each other:
\begin{equation}
\avnuq{p_T}{q}_{\delta}(\beta; q) =  \avnuq{p_T}{1}_{\delta}(q\beta; 2-1/q) , 
\end{equation}
where we use the relation $(e_q(x))^q = e_{2-1/q}(qx)$ for $q>0$ \cite{Umeno-prepri}.

The following integral \cite{Ishihara2015:BG} for $q>1$ appears in the average.
\begin{equation}
I(\rho, \xi=0) = \int_0^{\infty} dk \frac{k^{\rho}}{\left( 1+(q-1) \beta k \right)^{1/(q-1)}}  \qquad (q>1) . 
\label{def:integral:I}
\end{equation}
With this integral,  we have a simple expression of the average:
\begin{equation}
\avnuq{p_T}{1}_{\delta} (\beta; q) = \frac{I(\rho=2+\delta,\xi=0)}{I(\rho=1+\delta,\xi=0)}  . 
\end{equation}

We obtain the averages, $\avnuq{p_T}{1}_{\delta} (\beta; q)$ and $\avnuq{p_T}{q}_{\delta} (\beta; q)$, for $\delta =0, 1$:
\begin{subequations}
\begin{align}
&\avnuq{p_T}{1}_{\delta=0} (\beta; q) = \frac{2}{\beta} \frac{1}{(4-3q)} \qquad (1<q<\frac{4}{3}) ,\\
&\avnuq{p_T}{1}_{\delta=1} (\beta; q) = \frac{3}{\beta} \frac{1}{(5-4q)} \qquad (1<q<\frac{5}{4}) ,\\
&\avnuq{p_T}{q}_{\delta=0} (\beta; q) = \frac{2}{\beta} \frac{1}{(3-2q)} \qquad (1<q<\frac{3}{2}) ,\\
&\avnuq{p_T}{q}_{\delta=1} (\beta; q) = \frac{3}{\beta} \frac{1}{(4-3q)} \qquad (1<q<\frac{4}{3}) .
\end{align}
\end{subequations}

\subsection{Transverse momentum fluctuation under the Tsallis distribution}
The fluctuation has been measured with the $p_T$ fluctuation measure \cite{ALICE-Collaboration}.
The following measure $R(N)$ is calculated in this study:
\begin{subequations}
\begin{align}
&R(N) := \frac{Q(N)}{N(N-1)} ,\\
&Q(N) := \int d^2\vec{p}_T^{(1)} \int dy^{(1)} \int d^2\vec{p}_T^{(2)} \int dy^{(2)} \ \left( \frac{d^6N}{d\vec{p}_T^{(1) 2} dy^{(1)} d\vec{p}_T^{(2) 2}dy^{(2)}} \right)
\nonumber \\ & \qquad\qquad\qquad  \times
\left( p_T^{(1)} - \av{ p_T^{(1)}} \right)\left( p_T^{(2)} - \av{ p_T^{(2)}} \right) .  
\end{align}
\end{subequations}

We assume an exponential type correlation between two particles, which is frequently used as Bose-Einstein type correlation. 
That is
\begin{align}
& \left( \frac{d^6N}{d\vec{p}_T^{(1) 2} dy^{(1)} d\vec{p}_T^{(2) 2} dy^{(2)}} \right)
\nonumber \\ &\qquad\qquad
= \left( 1 + \lambda \exp \left( - a^2 \left| \vec{p}_T^{(1)} - \vec{p}_T^{(2)} \right|^2 \right) \right)
\left( \frac{d^3N}{d\vec{p}_T^{(1) 2} dy^{(1)} } \right) \left( \frac{d^3N}{d\vec{p}_T^{(2) 2} dy^{(2)} } \right)
,
\end{align}
where $\lambda$ is independent of $\vec{p}_T$ and $y$.
The parameter $\lambda$ is the correlation strength and the parameter $a$ is the correlation length.
With the definition of the average of $p_T$, we have 
\begin{align}
Q(N) &=  \lambda \int d^2\vec{p}_T^{(1)} \int dy^{(1)} \int d^2\vec{p}_T^{(2)} \int dy^{(2)} \ 
\exp \left( - a^2 \left| \vec{p}_T^{(1)} - \vec{p}_T^{(2)} \right|^2 \right) 
\nonumber \\
& \quad\quad\times
\left( \frac{d^3N}{d\vec{p}_T^{(1) 2} dy^{(1)} } \right) \left( \frac{d^3N}{d\vec{p}_T^{(2) 2} dy^{(2)} } \right) 
\left( p_T^{(1)} - \av{ p_T^{(1)}} \right)\left( p_T^{(2)} - \av{ p_T^{(2)}} \right)  
.
\label{eqn:temp:Q}
\end{align}

The $p_T$ fluctuation measure for charged hadrons $R_{\mathrm{c}}$ is evidently described with the number of partons $\nudelta{N}$ for $\nudeltac{N} \gg 1$,  
because $\nudeltac{N} ( \nudeltac{N} -1 ) $ is approximately $\left( \nudeltac{N} \right)^2=\gamma^{-2} \left( \nudelta{N} \right)^2$:  
we cancel out the common factors,  $\gamma^{-2}$, in the numerator and the denominator of the $p_T$ fluctuation measure:
\begin{equation}
R_{\mathrm{c}} \equiv R(\nudeltac{N}) 
\sim \frac{Q(\nudeltac{N})}{\left( \nudeltac{N} \right)^2}  
= \frac{Q(\nudelta{N})}{ \left( \nudelta{N} \right)^2} \sim R(\nudelta{N}) 
.
\end{equation}

We use Eqs.~\eqref{eqn:bjorken} and \eqref{eqn:used-dist} to calculate Eq.~\eqref{eqn:temp:Q}.
After the integration with respect to angle, we have
\begin{align}
\nudelta{Q}(\nudelta{N}) 
&
= \lambda \left( \nudelta{C} \right)^2 
\int dp_T^{(1)} \int dp_T^{(2)} 
\exp \left( - a^2 \left[ \left( p_{T}^{(1)} \right)^2 + \left( p_{T}^{(2)} \right)^2 \right] \right)
    \nonumber \\ & \quad \times
\left( p_{T}^{(1)} p_{T}^{(2)} \right)^{1+\delta} I_0 \left( 2 a^2  p_{T}^{(1)} p_{T}^{(2)} \right) 
\left( \frac{1}{e_q\left( \beta  p_{T}^{(1)} \right)} \right)^{\nu} \left( \frac{1}{e_q\left( \beta  p_{T}^{(2)} \right)} \right)^{\nu}
    \nonumber \\ & \quad \times
\left( p_T^{(1)} - \avnu{ p_T^{(1)}}_{\delta} \right)\left( p_T^{(2)} - \avnu{ p_T^{(2)}}_{\delta} \right) 
, 
\end{align}
where we attach the superscript $(\nu)$ and the subscript $\delta$ to $Q$, 
and the function $I_0(t)$ is a modified Bessel function. 
The expansion of $I_0(t)$ \cite{Abramowitz, Moriguchi} is ${\displaystyle \sum_{n=0} \frac{(t/2)^{2n}}{(n!)^2}}$. 
We obtain the expansion of $\nudelta{Q}$  with the expansion of $I_0(t)$ 
with $\eta = \beta/a$ as follows:
\begin{subequations}
\begin{align}
&
\nudelta{Q}(\nudelta{N}) = \frac{\lambda \left( \nudelta{C} \right)^2}{a^{6+2\delta}} 
\sum_{n=0} \frac{1}{(n!)^2} \left[ \tilde{M}^{(\nu)}_{2n+2+\delta}(\eta; q)  - \avnu{\tilde{p}_T}_{\delta} \tilde{M}^{(\nu)}_{2n+1+\delta}(\eta; q) \right]^2 
, \\ 
&
\tilde{M}^{(\nu)}_j(\eta; q)  := \int_0^{\infty} \ dt\ e^{-t^2} t^j \left( \frac{1}{e_q(\eta t)} \right)^{\nu}   
, \\ 
&
\avnu{\tilde{p}_T}_{\delta} := a \avnu{p_T}_{\delta}
.
\end{align}
\end{subequations}

The number $\nudelta{N}$ is easily calculated: 
\begin{equation}
\nudelta{N} = \nudelta{C} \int_0^{\infty} dp_T \left(p_T\right)^{1+\delta}  \left( \frac{1}{e_q(\eta t)} \right)^{\nu} 
.
\end{equation}
The number $\nudelta{N}$ is given as follows:
\begin{subequations}
\begin{align}
\nudelta{N}(\beta; q) &= \left( \frac{\nudelta{C}}{\beta^{2+\delta}} \right)  \nudelta{\tilde{N}}(q) , \label{eqn:N-tildeN}\\
& \tilde{N}_{\delta=0}^{(1)}(q) = \frac{1}{(3-2q)(2-q)} , \\
& \tilde{N}_{\delta=1}^{(1)}(q) = \frac{2}{(4-3q)(3-2q)(2-q)} , \\
& \tilde{N}_{\delta=0}^{(q)}(q) = \frac{1}{(2-q)} , \\
& \tilde{N}_{\delta=1}^{(q)}(q) = \frac{2}{(3-2q)(2-q)} . 
\end{align}
\end{subequations}

The $p_T$ fluctuation measure $\nudelta{R} \equiv R(\nudelta{N})$ for $\nudelta{N} \gg 1$ is obtained with these expressions:
\begin{align}
\nudelta{R} = 
\frac{\lambda}{a^2} \frac{ \eta^{4+2\delta}}{(\nudelta{\tilde{N}}(q) )^2} 
\sum_{n=0} \frac{1}{(n!)^2} \left[ \tilde{M}^{(\nu)}_{2n+2+\delta}(\eta; q)  - \avnu{\tilde{p}_T}_{\delta} \tilde{M}^{(\nu)}_{2n+1+\delta}(\eta; q) \right]^2 
. 
\end{align}
The quantity $\sqrt{R_{\mathrm{c},\delta}^{(\nu)}}/\av{p_T}_{\mathrm{c}, \delta}^{(\nu)}$ is frequently used in the analysis. 
We obtain approximately the quantity $\sqrt{R_{\mathrm{c},\delta}^{(\nu)}}/\av{p_T}_{\mathrm{c}, \delta}^{(\nu)}$ for charged hadrons: 
\begin{align}
&
\frac{\sqrt{R_{\mathrm{c},\delta}^{(\nu)}}}{\av{p_T}_{\mathrm{c}, \delta}^{(\nu)}}
\sim 
\frac{\sqrt{\nudelta{R}}}{\avnu{p_T}_\delta}
\nonumber \\ 
& \quad \sim  
\frac{\sqrt{\lambda} \eta^{2+\delta}}{ \avnu{\tilde{p}_T}_{\delta}  \nudelta{\tilde{N}}(q) } 
\sqrt{\sum_{n=0} \frac{1}{(n!)^2} \left[ \tilde{M}^{(\nu)}_{2n+2+\delta}(\eta; q)  - \avnu{\tilde{p}_T}_{\delta} \tilde{M}^{(\nu)}_{2n+1+\delta}(\eta; q) \right]^2 }
.
\label{eqn:final}
\end{align}
We note that the right-hand side of Eq.~\eqref{eqn:final} depends on the combination of the variables, $a$ and $\beta$:
Eq.~\eqref{eqn:final} depends on $\eta=\beta/a$.


\section{Numerical Estimation}
\label{sec:numerical}
The following quantity $\nudelta{U}(\eta, q)$ was calculated numerically for various $\eta$ and $q$ 
to study the effects of the form of the distribution 
and 
to study the differences between the fluctuation for the expectation value used in the BG statistics
and the fluctuation for the expectation value used in the Tsallis statistics: 
\begin{align}
\nudelta{U}(\eta, q) := \left( 1/\sqrt{\lambda} \right) \left(\sqrt{\nudelta{R}}/\avnu{p_T}_\delta\right) . 
\label{eqn:U:R-pt}
\end{align}

The values of the parameters were adopted by referring the results in high energy collisions.
The value of $q-1>0$ is small in many cases \cite{Cleymans-conf}. 
Therefore, the range of $q$ was $1.0 \le q \le 1.1$ in the numerical calculations. 
The value of $\eta$ is also required to calculate $\nudelta{U}(\eta,q)$.  
The range of $\eta$ was $4 \le \eta \le 8$ in the numerical calculations.  
In Ref.~\cite{our},
the temperature $\beta^{-1}$ equals $\kappa Q_{\mathrm{sat}}$, where $Q_{\mathrm{sat}}$  is saturation momentum.
The parameter $a^{-1}$ equals $\sqrt{\sigma} Q_{\mathrm{sat}}$.
Therefore, $\eta$ is rewritten with the model parameters:
\begin{equation}
\eta = \frac{\beta}{a} = \frac{\sqrt{\sigma}}{\kappa} 
.
\label{eqn:eta-other-model-parameters}
\end{equation}
The value $\eta$ estimated from the parameters for the fits of all data points (fit-A) in Ref.~\cite{our} is in the range of $4$ to $8$. 

\begin{figure}
\begin{center}
\includegraphics[width=0.45\textwidth]{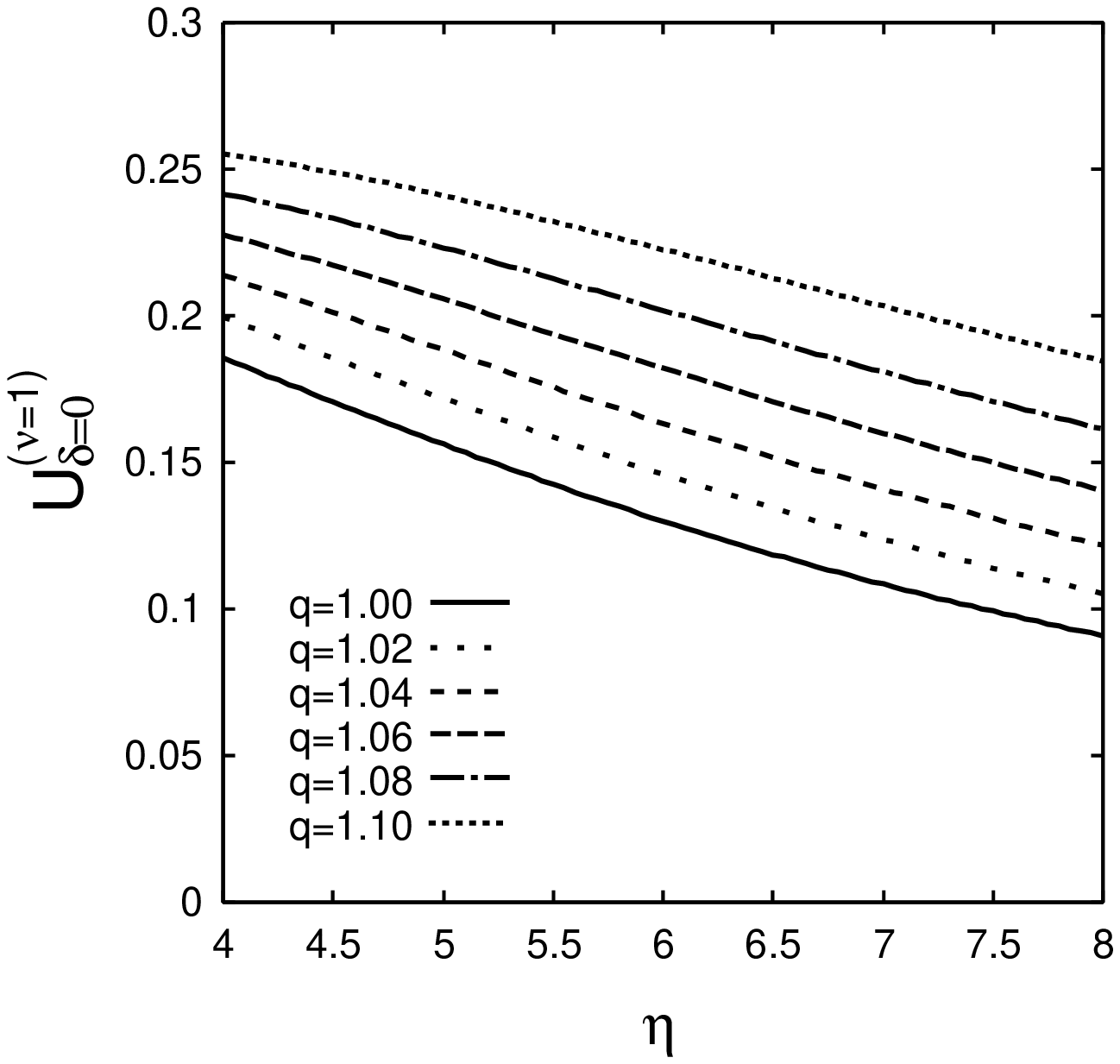}
\end{center}
\caption{The quantity  $U^{(\nu=1)}_{\delta=0}(\eta, q)$ as a function of $\eta$ for various $q$.}
\label{fig:BG:GS}
\end{figure}

\begin{figure}
\begin{center}
\includegraphics[width=0.45\textwidth]{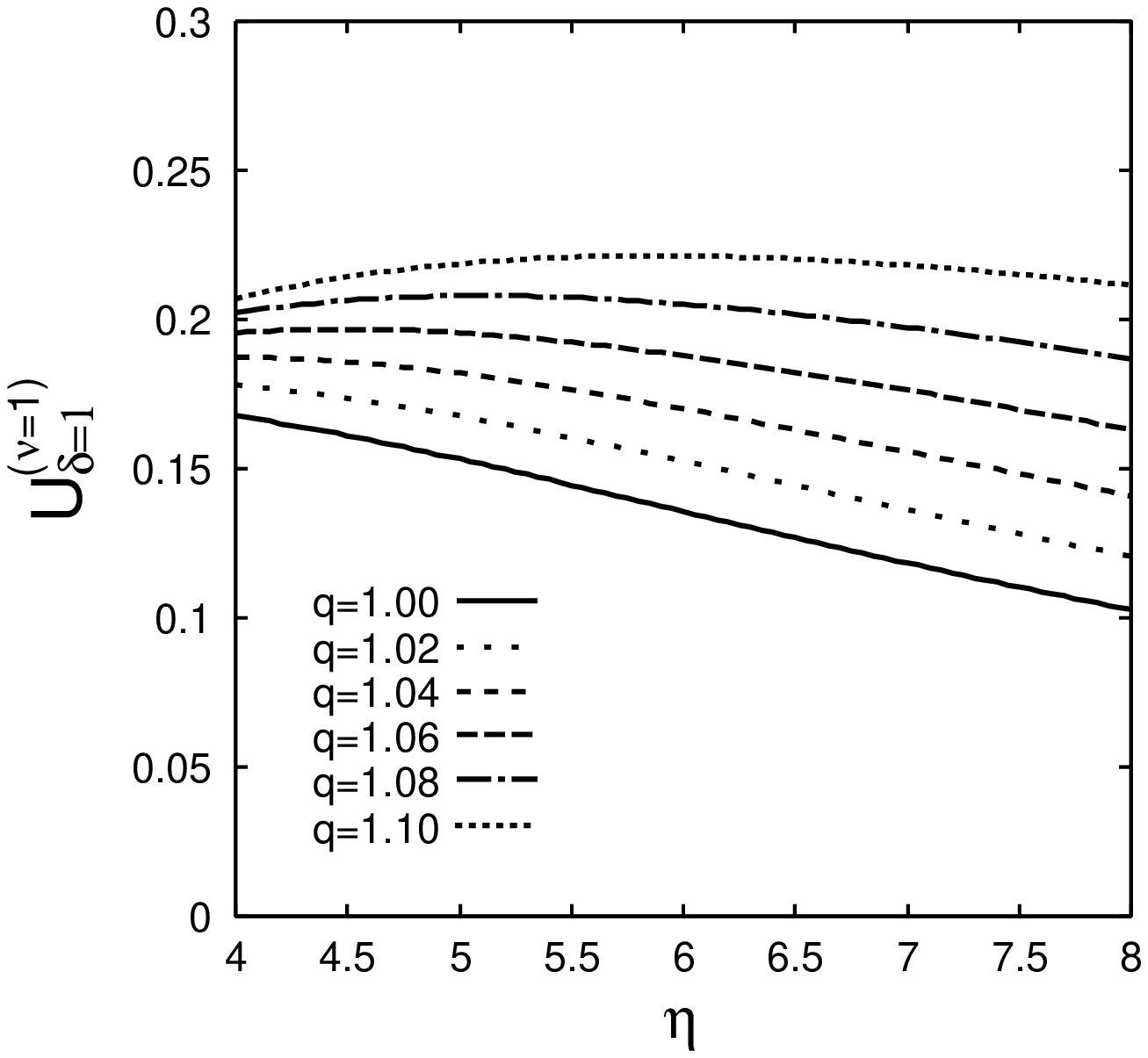}
\end{center}
\caption{The quantity  $U^{(\nu=1)}_{\delta=1}(\eta, q)$ as a function of $\eta$ for various $q$.}
\label{fig:BG:Bj}
\end{figure}

Figure~\ref{fig:BG:GS} and Figure~\ref{fig:BG:Bj} show 
$U^{(\nu=1)}_{\delta=0}(\eta, q)$ and  $U^{(\nu=1)}_{\delta=1}(\eta, q)$ as functions of $\eta$ for various $q$.
The $q$ dependences are similar in both the figures:
the value $U^{(\nu=1)}_{\delta=0}(\eta, q)$ is larger than $U^{(\nu=1)}_{\delta=0}(\eta, q')$ 
and the value $U^{(\nu=1)}_{\delta=1}(\eta, q)$ is larger than $U^{(\nu=1)}_{\delta=1}(\eta, q')$,
 when $q$ is larger than $q'$. 
In contrast, 
the $\eta$ dependence of $U^{(\nu=1)}_{\delta=1}(\eta, q)$ is different from that of  $U^{(\nu=1)}_{\delta=0}(\eta, q)$.
In Figure~\ref{fig:BG:GS}, the value $U^{(\nu=1)}_{\delta=0}(\eta, q)$ decreases as $\eta$ increase in the range of $4 \le \eta \le 8$. 
In Figure~\ref{fig:BG:Bj}, in the range of $4 \le \eta \le 8$,
the value $U^{(\nu=1)}_{\delta=1}(\eta, q)$ for small $q$ decreases as $\eta$ increases, 
and  $U^{(\nu=1)}_{\delta=1}(\eta, q)$ for large $q$ has an extremum. 


\begin{figure}
\begin{center}
\includegraphics[width=0.45\textwidth]{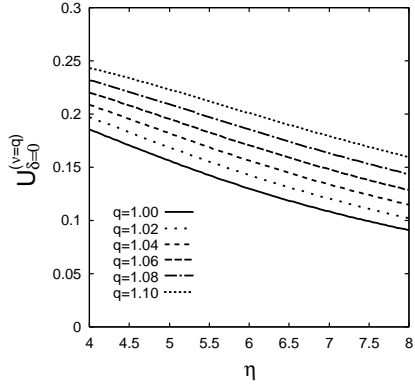}
\end{center}
\caption{The quantity  $U^{(\nu=q)}_{\delta=0}(\eta, q)$ as a function of $\eta$ for various $q$.}
\label{fig:Tsallis:GS}
\end{figure}

\begin{figure}
\begin{center}
\includegraphics[width=0.45\textwidth]{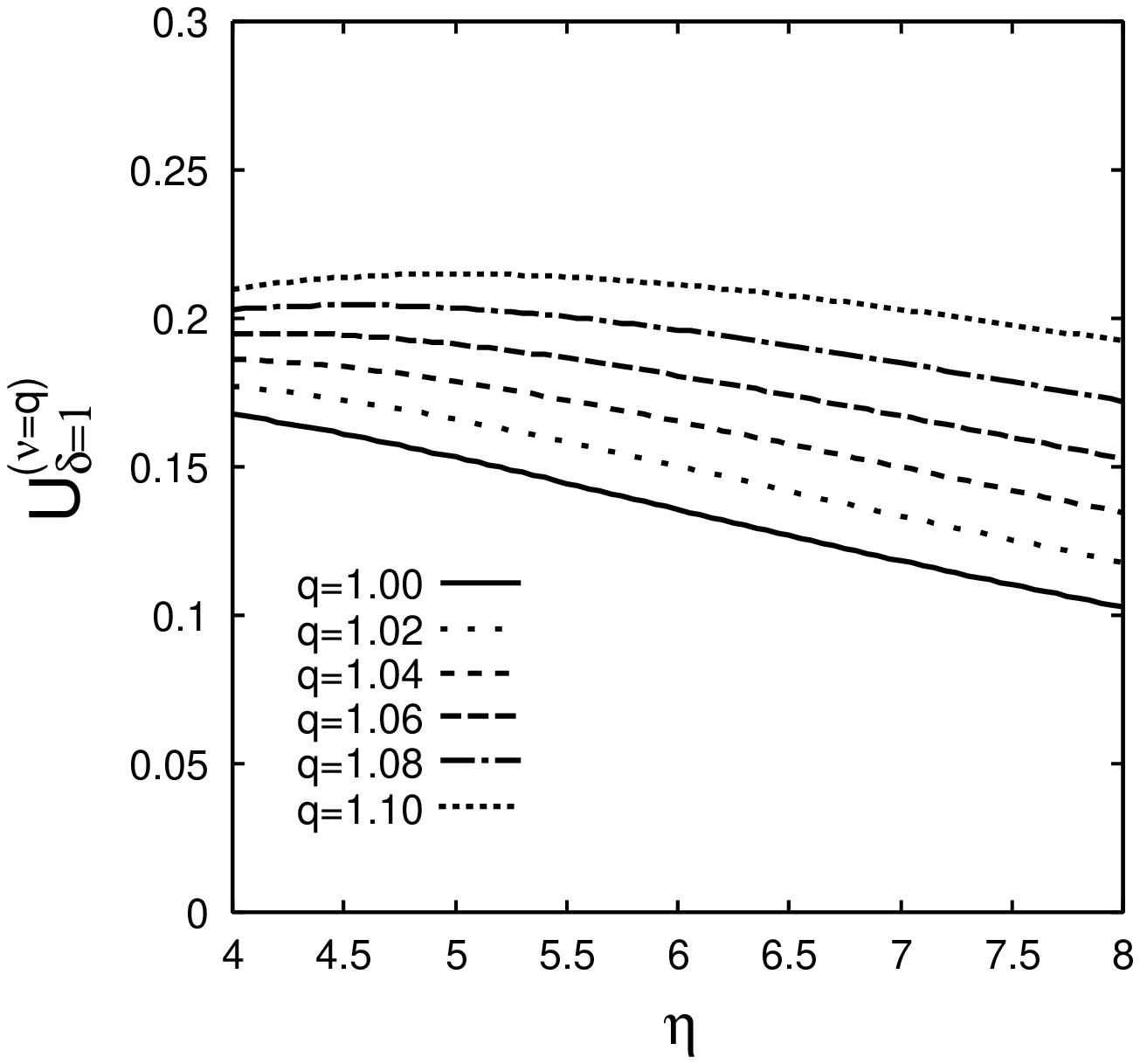}
\end{center}
\caption{The quantity  $U^{(\nu=q)}_{\delta=1}(\eta, q)$ as a function of $\eta$ for various $q$.}
\label{fig:Tsallis:Bj}
\end{figure}

Figure~\ref{fig:Tsallis:GS} and  Figure~\ref{fig:Tsallis:Bj} show 
$U^{(\nu=q)}_{\delta=0}(\eta, q)$ and $U^{(\nu=q)}_{\delta=1}(\eta, q)$ as functions of $\eta$ for various $q$.
The behavior  of $U^{(\nu=q)}_{\delta=0}(\eta, q)$ in Figure~\ref{fig:Tsallis:GS} is similar to that of  $U^{(\nu=1)}_{\delta=0}(\eta, q)$ in Figure~\ref{fig:BG:GS},
and 
the behavior  of $U^{(\nu=q)}_{\delta=1}(\eta, q)$ in Figure~\ref{fig:Tsallis:Bj} is also similar to that of  $U^{(\nu=1)}_{\delta=1}(\eta, q)$ in Figure~\ref{fig:BG:Bj}.

\begin{figure}
\begin{center}
\includegraphics[width=0.45\textwidth]{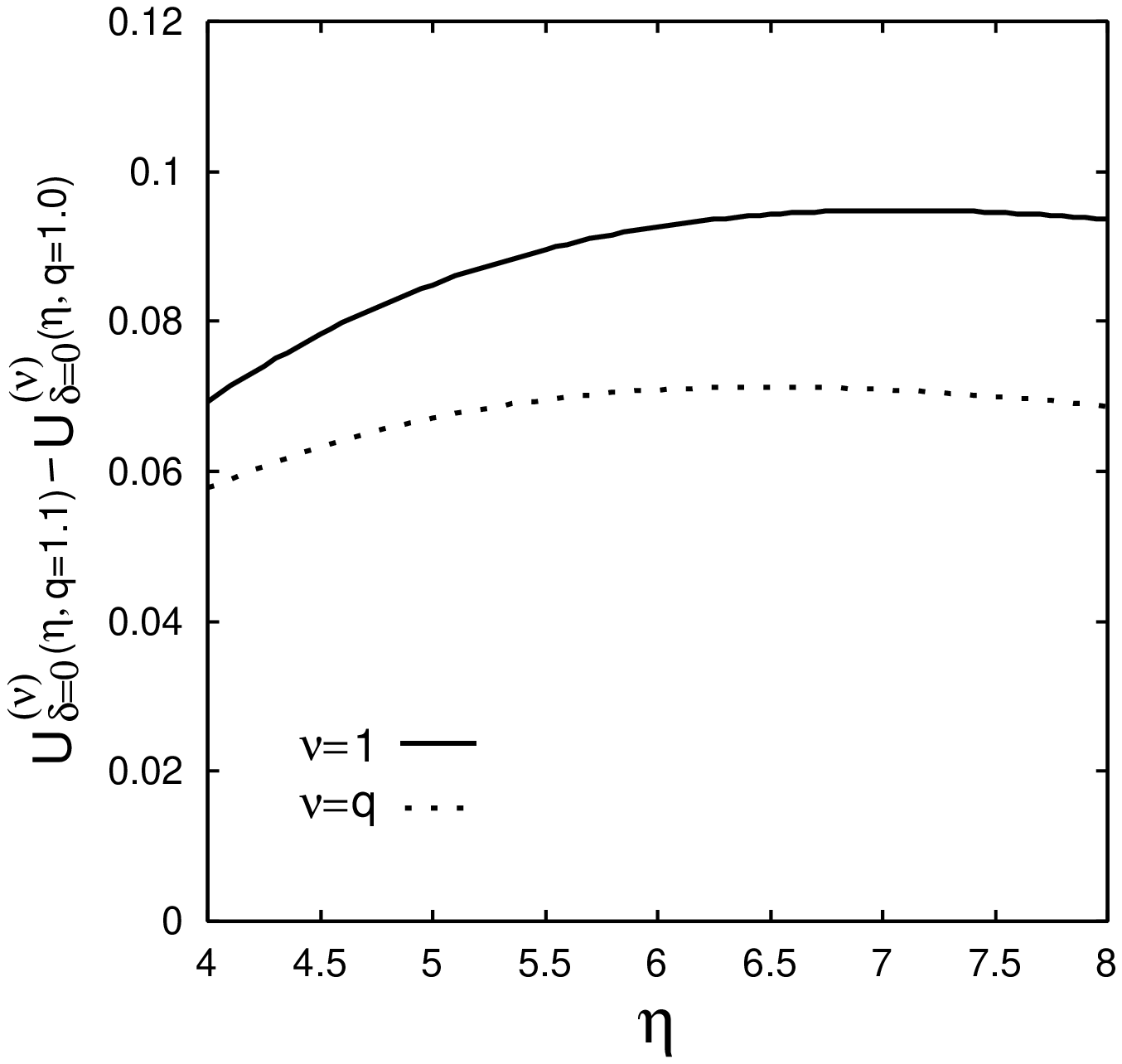}
\includegraphics[width=0.45\textwidth]{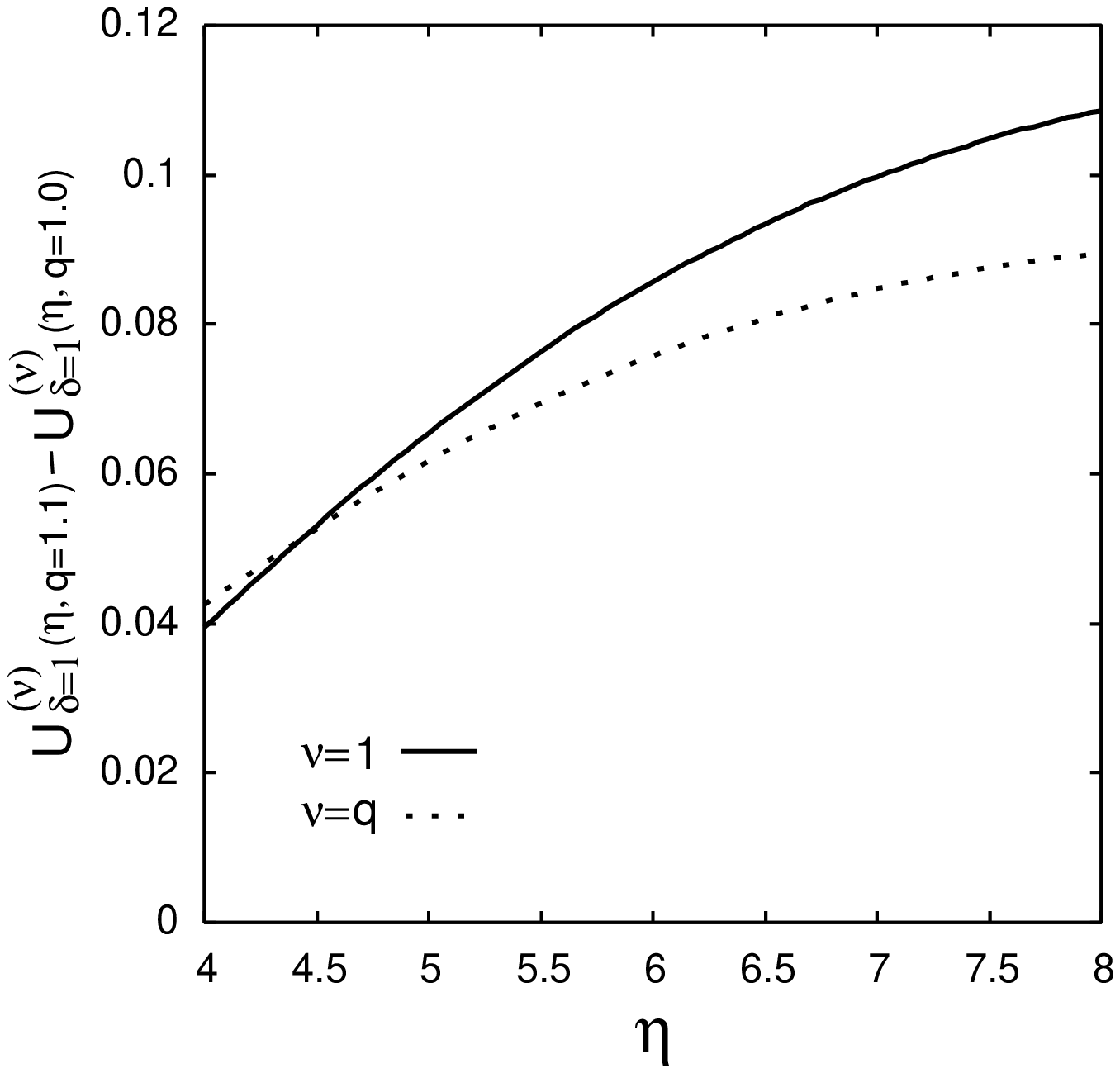} 
\end{center}
\caption{
The quantity $U_{\delta}^{(\nu)}(\eta, q=1.1) - U_{\delta}^{(\nu)}(\eta, q=1.0)$ in the range of $4 \le \eta \le 8$.
The left panel shows  the quantity for $\delta=0$, and 
the right panel shows  the quantity for $\delta=1$.
} 
\label{fig:diff-q-in-nu}
\end{figure}

Figure~\ref{fig:diff-q-in-nu} shows the quantity $U_{\delta}^{(\nu)}(\eta,q=1.1)  - U_{\delta}^{(\nu)}(\eta,q=1.0)$ 
in order to study the $q$-dependence of $U_{\delta}^{(\nu)}(\eta,q)$.
As a function of $q$, the variation of $U_{\delta=0}^{(\nu=1)}$ is larger than that of $U_{\delta=0}^{(\nu=q)}$ from the left panel of Fig.~\ref{fig:diff-q-in-nu}:
for $\delta=0$, the $q$ dependence for the expectation value used in the BG statistics is larger than 
that for the expectation value used in the Tsallis nonextensive statistics in the range of $4 \le \eta \le 8$. 
As a function of $q$, the variation of $U_{\delta=1}^{(\nu=1)}$ is also larger than that of $U_{\delta=1}^{(\nu=q)}$ except for the vicinity of $\eta=4.5$
from the right panel of Fig.~\ref{fig:diff-q-in-nu}.
For $\delta=1$, the variation of $U_{\delta=1}^{(\nu=1)}$ is close to that of $U_{\delta=1}^{(\nu=q)}$ in the vicinity of $\eta=4.5$.
The variation of $U_{\delta}^{(\nu=1)}$ is larger than that of $U_{\delta}^{(\nu=q)}$ in the wide range of $\eta$, as shown in Fig.~\ref{fig:diff-q-in-nu}.


\begin{figure}
\begin{center}
\includegraphics[width=0.45\textwidth]{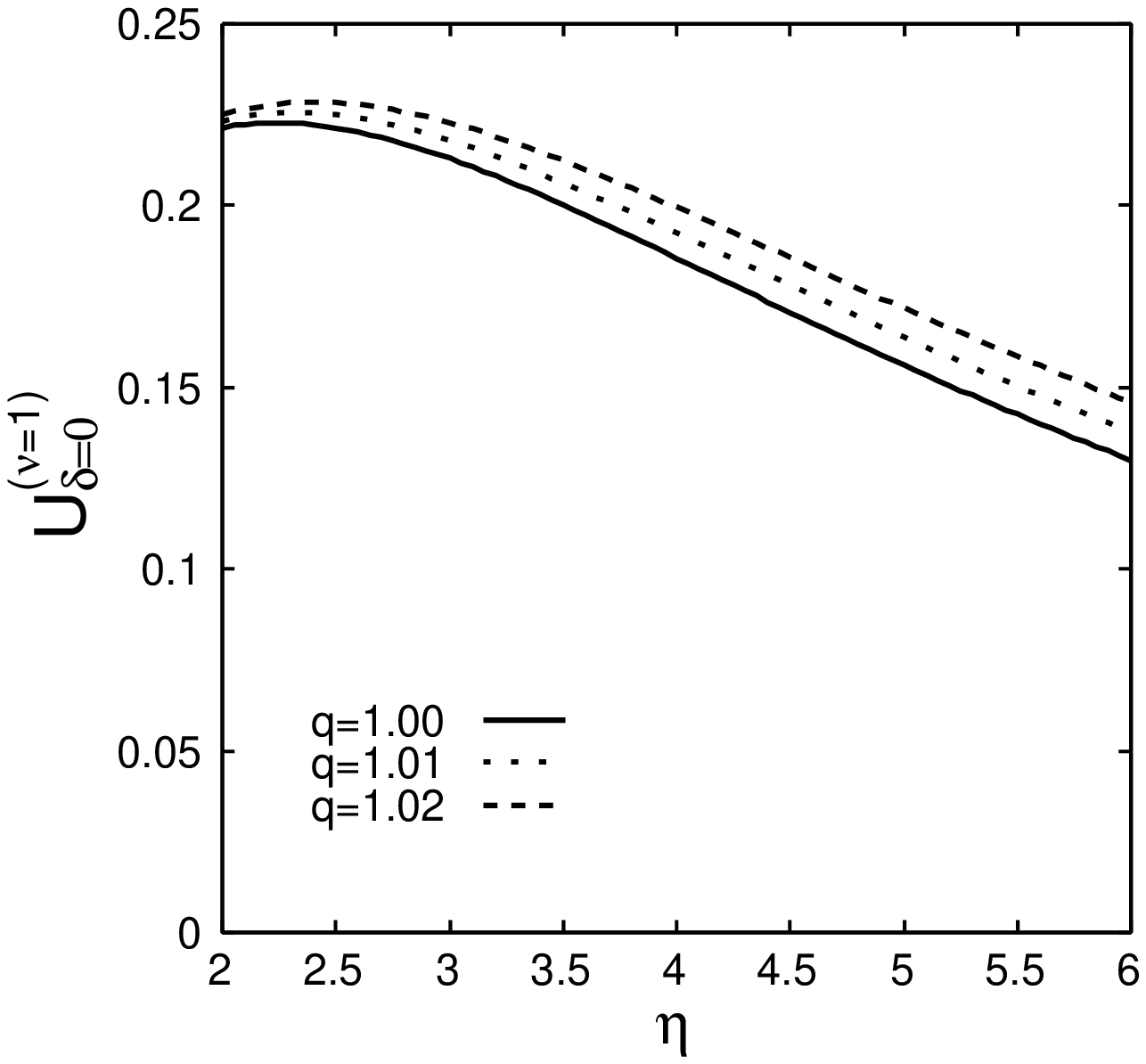}
\includegraphics[width=0.45\textwidth]{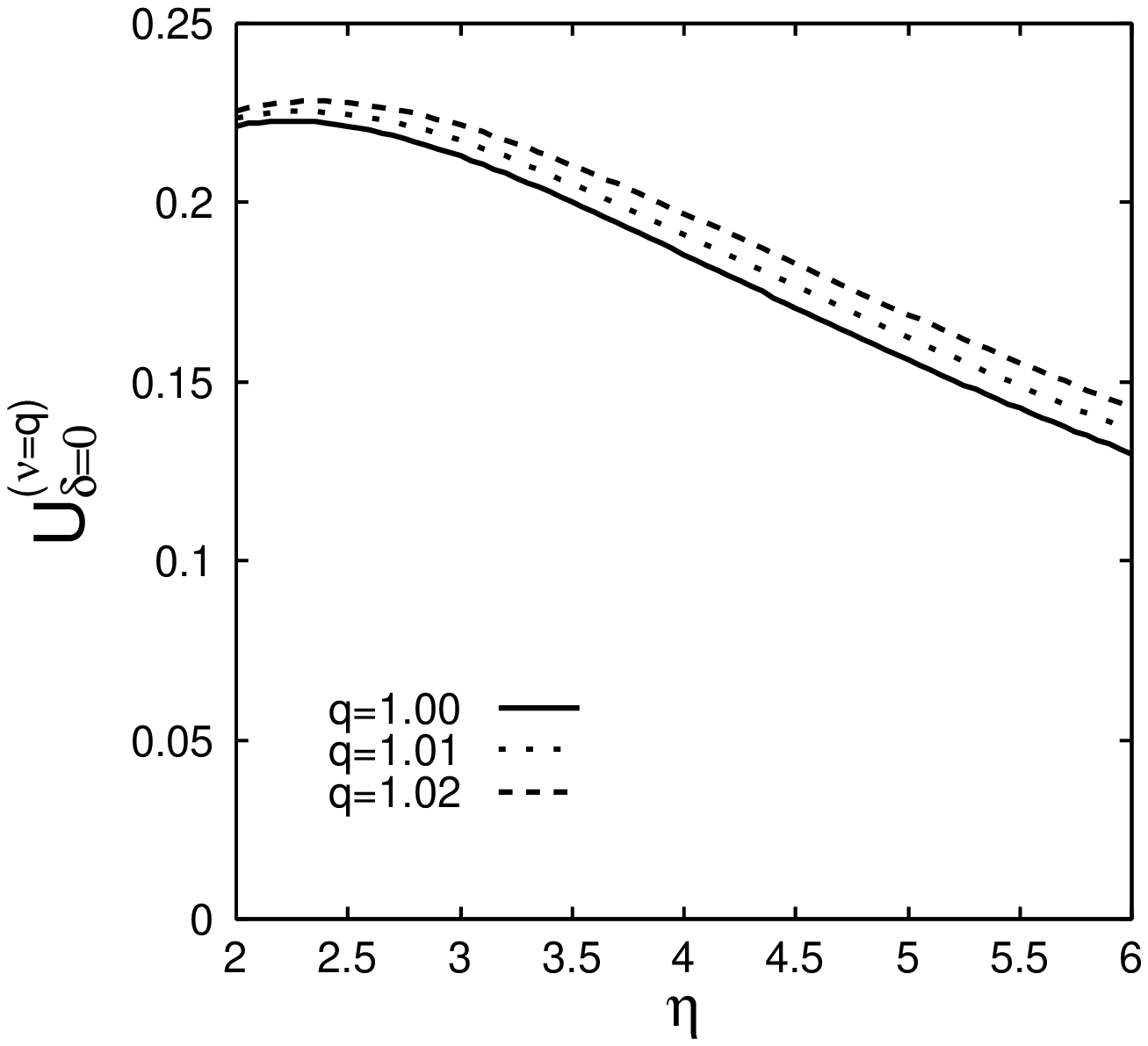}
\end{center}
\caption{The quantity $U_{\delta}^{(\nu)}$ in the range of $2 < \eta \le 6$ for $q=1.00, 1.01$, and $1.02$. 
The quantity $U^{(\nu=1)}_{\delta=0}(\eta, q)$ is shown in the left panel,  
and $U^{(\nu=q)}_{\delta=0}(\eta, q)$ is shown in the right panel.
} 
\label{fig:GSwide-range}
\end{figure}

Apart from the values of the parameters in high energy collisions, 
we calculated the quantity $U_{\delta=0}^{(\nu)}$ numerically to study the $\eta$ dependence of $U_{\delta=0}^{(\nu)}$.
The quantity $U_{\delta=0}^{(\nu)}$ was calculated in the range of $2 < \eta \le 6$ for $q = 1.00, 1.01,$ and $1.02$.
Figure~\ref{fig:GSwide-range} shows $U_{\delta=0}^{(\nu=1)}$ and $U_{\delta=0}^{(\nu=q)}$.
The quantity $U_{\delta=0}^{(\nu)}$ has an extremum in both the cases. 
These numerical results indicate that the quantity $U_{\delta=0}^{(\nu)}$ is not a monotonically decreasing function of $\eta$ in the wide range of $\eta$.

\begin{figure}
\begin{center}
\includegraphics[width=0.45\textwidth]{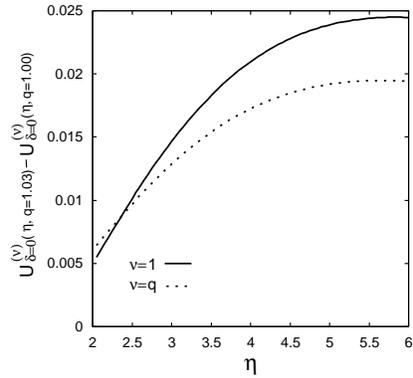}
\end{center}
\caption{The quantity $U_{\delta=0}^{(\nu)}(\eta, q=1.03) - U_{\delta=0}^{(\nu)}(\eta, q=1.00)$ in the range of $2 < \eta \le 6$}
\label{fig:GS:diff-small-q}
\end{figure}

Figure~\ref{fig:GS:diff-small-q} shows the quantity  $U_{\delta=0}^{(\nu)}(\eta, q=1.03) - U_{\delta=0}^{(\nu)}(\eta, q=1.00)$. 
Figure~\ref{fig:GS:diff-small-q} is similar to the right panel of Fig.~\ref{fig:diff-q-in-nu}.
The effects in the case of the expectation value used in the BG statistics is slightly larger than 
that in the case of the expectation value used  in the Tsallis  nonextensive statistics in the wide range of $\eta$, as shown in Fig.~\ref{fig:GS:diff-small-q}.

\section{Discussion and Conclusion}
We studied the transverse momentum fluctuation of the emitted particles in high energy collisions, assuming the parton-hadron duality. 
The average of the absolute value of the transverse momentum and the fluctuation were calculated 
under the Tsallis distribution with the entropic parameter $q$:
the distribution appears in the studies of high energy collisions.
These quantities were calculated for the expectation value used in the Boltzmann-Gibbs  statistics and 
for the expectation value used in the Tsallis nonextensive statistics.

We introduce the quantity $\nudelta{U}$, which is related to the $p_T$ fluctuation measure, 
to study the parameter dependence of the fluctuation numerically, 
where $\delta$ distinguishes the power of $p_T$ and $\nu$ distinguishes the definition of the expectation value.
The quantity $\nudelta{U}$ has two parameters, $q$ and $\eta$:  
the parameter $\eta$ is the ratio of the inverse temperature $\beta$ to the correlation length $a$,
where $a$ is introduced in the Bose-Einstein type correlation between partons. 
The values of the fluctuations were numerically estimated.

We obtained the following results:
(1) the quantity $\nudelta{U}$ as a function of $\eta$ depends on the form of the distribution and 
depends weakly on the definition of the expectation value used in the statistics,
(2) the quantity $\nudelta{U}$ increases as $q$ increases, 
and 
(3) the variation of the quantity $\nudelta{U}$ as a function of $q$ 
for the expectation value used in the BG statistics is larger than that for the expectation value used in the Tsallis nonextensive statistics 
in the wide range of $\eta$.
In addition, the quantity $\nudelta{U}$ is not a monotonically decreasing function of $\eta$ in the wide range of $\eta$.

These results indicate the following points.
The result (1) indicates that the $\eta$-dependence of the fluctuation depends on 
the form of the transverse momentum distribution function.
As shown in the result (2), the value of the fluctuation is an increasing function of $q$. 
The $q$-dependence of the fluctuation should come from the tail of the Tsallis distribution.
The result (3) indicates that 
the effects of the Tsallis distribution for the expectation value used in the BG statistics are larger than 
those for the expectation value used in the Tsallis nonextensive statistics in the wide range of $\eta$. 

We finally discuss the quantities estimated from $U_{\delta}^{(\nu)}(\eta,q)$ in high energy collisions.
The entropic parameter $q$ and the temperature $T$ are obtained by fitting the distribution, as researchers studied. 
The values, $\delta$ and $\nu$, will be determined by the fits. 
The remaining parameters are the correlation strength $\lambda$ and the correlation length $a$.
In the case that $\lambda$ and $a$ are assumed to be constant,
the quantities $\lambda$ and $a$ can be obtained by fitting $\sqrt{\lambda} U_{\delta}^{(\nu)}(\eta,q)$ to the observed fluctuation measure. 
In the case that $a$ is given,
the quantity $\sqrt{\lambda}$ can be obtained as the ratio of the observed fluctuation measure to $U(\eta, q)$: 
for example, as for $a$, the flux tube size is a candidate of the length $a$.
The quantity $U_{\delta}^{(\nu)}(\eta,q)$ is related to the physical quantities under some assumptions.

The findings in this study are useful to discuss the fluctuation 
when the distribution is a power-like distribution such as a Tsallis distribution.  
We hope that this work is helpful to understand the phenomena at high energies.

\section*{Acknowledgments}
The author thanks Dr.~T.~Osada for helpful discussions.


\appendix 
\section{Integral $I(\rho,\xi=0)$}
The integral $I(\rho,\xi)$ defined in Eq.~\eqref{def:integral:I} appears in the calculation.  
We give the results of the integrals with $\xi=0$ for $\rho=1, 2$, and $3$. 
The integration is easily performed by changing of variables \cite{Ishihara2015:BG}: 
\begin{equation}
I(\rho,\xi) = \frac{1}{\beta^{\rho+1} (q-1)^{\rho}} \int_0^1 dy \ \frac{y^{1-q} \left( y^{1-q} - 1\right)^{\rho}}{1+\xi y}
. 
\end{equation}
The results are given as follows: 
\begin{subequations}
\begin{align}
&I(\rho=1,\xi=0) = \frac{1}{\beta^2} \frac{1}{(3-2q)(2-q)} &\quad (1<q<\frac{3}{2}) , \\
&I(\rho=2,\xi=0) = \frac{2}{\beta^3} \frac{1}{(4-3q)(3-2q)(2-q)} &\quad (1<q<\frac{4}{3}) , \\
&I(\rho=3,\xi=0) = \frac{6}{\beta^4} \frac{1}{(5-4q)(4-3q)(3-2q)(2-q)} &\quad (1<q<\frac{5}{4})  . 
\end{align}
\end{subequations}


\begin{thebibliography}{20}

\bibitem{Alberico2000} W.~M.~Alberico, L.~Lavagno, and P.~Quarati, 
Eur.~Phys.~J.~C \textbf{12}, 499 (2000). 

\bibitem{Wilk2007} G.~Wilk, 
Brazillian Journal of Physics \textbf{37}, 714 (2007). 

\bibitem{Osada2008} T.~Osada and G.~Wilk, 
Phys.~Rev.~C \textbf{77}, 044903 (2008).


\bibitem{Alberico2009} W.~M.~Alberico, L.~Lavagno,
Eur.~Phys.~J.~A \textbf{40}, 313 (2009).

\bibitem{Cleymans2012} J.~Cleymans and D.~Worku,  
J.~Phys.~G: Nucl.~Part.~Phys.~\textbf{39}, 025006 (2012).


\bibitem{Marques2015} L.~Marques, J.~Cleymans, and A.~Deppman, 
Phys.~Rev.~D \textbf{91}, 054025 (2015).


\bibitem{Cleymans2017} A.~Khuntia, S.~Tripathy, R.~Sahoo, and J.~Cleymans, 
arXiv:1702.06885. 



\bibitem{GS2014} C.~K.-Boesing and L.~McLerran, 
Phys.~Lett.~B \textbf{734}, 282 (2014). 


\bibitem{GS2015} L.~McLerran and M.~Praszalowicz, 
Phys.~Lett.~B \textbf{741}, 246 (2015). 


\bibitem{Zheng2016}  H.~Zheng and L.~Zhu,
Advances in High Energy Physics, Vol.~\textbf{2016}, 9632126 (2016). 


\bibitem{Lao2016} H.-L.~Lao, F.-H.~Liu, and R.~A.~Lacey,
Eur.~Phys.~J.~A \textbf{53}, 44 (2017).


\bibitem{our} T.~Osada and M.~Ishihara, 
Journal of Physics G: Nuclear and Particle Physics, in press.  
The digital object identifier of the accepted manuscript is 10.1088/1361-6471/aa9208.



\bibitem{tsallis1998} C.~Tsallis, R.~S.~Mendes, and A.~R.~Plastino, 
Physica~A \textbf{261}, 534 (1998).


\bibitem{lavagno2002} A.~Lavagno, 
Phys.~Lett.~A \textbf{301}, 13 (2002).

\bibitem{Ishihara2015:BG} M.~Ishihara, 
Int.~J.~Mod.~Phys.~E \textbf{24}, 1550085 (2015). 


\bibitem{Ishihara2016:Tsallis} M.~Ishihara, 
Int.~J.~Mod.~Phys.~E \textbf{25}, 1650066 (2016). 


\bibitem{Lappi:conf:2011} T.~Lappi, 
J.~of~Physics: Conference Series \textbf{270}, 012055 (2011).


\bibitem{Bloom1971}  
E.~D.~Bloom and F.~J.~Gilman, 
Phys.~Rev.~D \textbf{4}, 2901 (1971).


\bibitem{Steffens2004} 
F.~M.~Steffens and K.~Tsushima, 
Phys.~Rev.~D \textbf{70}, 094040 (2004). 


\bibitem{Boris1996}
B.~Blok and T.~Mannel, 
Mod.~Phys.~Lett.~A \textbf{11}, 1263 (1996).



\bibitem{Kharzeev} 
D.~Kharzeev, E.~Levin, and L.~McLerran, 
Phys.~Lett.~B \textbf{561}, 93 (2003).



\bibitem{Conroy2010} J.~M.~Conroy, H.~G.~Miller, and A.~R.~Plastino, 
Phys.~Lett.~A \textbf{374}, 4581 (2010).


\bibitem{Umeno-prepri} K.~Umeno and  A.~Sato, 
arXiv:1205.1690.


\bibitem{ALICE-Collaboration} ALICE Collaboration, 
Eur.~Phys.~J.~C \textbf{74}:3077 (2014).


\bibitem{Abramowitz} M.~Abramowitz and I.~A.~Stegun, 
\textit{Handbook of Mathematical Functions with Formulas, Graphs, and Mathematical Tables} (Dover Publications, Inc., New York, 1965) .

\bibitem{Moriguchi} S.~Moriguchi, K.~Udagawa, and S.~Hitotsumatsu, 
\textit{Iwanami Suugaku koushiki III (Iwanami Mathematical Formulas III) } (Iwanami Shoten, Tokyo, 1987),  [in Japanese].


\bibitem{Cleymans-conf} J.~Cleymans, M.~D.~Azmi, A.~S.~Parvan, and O.~V.~Teryaev,
EPJ Web of Conferences \textbf{137}, 11004 (2017), 

\end{thebibliography}
\end{document}